\documentclass[conference]{IEEEtran}
\IEEEoverridecommandlockouts
\usepackage{cite}
\usepackage{amsmath,amssymb,amsfonts}
\usepackage{algorithmic}
\usepackage{graphicx}
\usepackage{float}
\usepackage{subfigure}
\usepackage{textcomp}
\usepackage{xcolor}
\def\BibTeX{{\rm B\kern-.05em{\sc i\kern-.025em b}\kern-.08em
    T\kern-.1667em\lower.7ex\hbox{E}\kern-.125emX}}
\begin{document}

\title{PSCNN: A 885.86 TOPS/W Programmable SRAM-based Computing-In-Memory Processor for Keyword Spotting}

\author{\IEEEauthorblockN{Shu-Hung Kuo, and Tian-Sheuan Chang, \textit{Senior Member, IEEE}}
% \IEEEauthorblockA{Institute of Electronics, National Yang Ming Chiao Tung Universoty, Taiwan}
% %\IEEEauthorblockA{\textit{, National Chiao Tung University}, Taiwan}
% }
\IEEEauthorblockA{\textit{Institute of Electronics, National Yang Ming Chiao Tung University,} \\
Hsinchu, Taiwan \\
}
}
\maketitle

\begin{abstract}
Computing-in-memory (CIM) has attracted significant attentions in recent years due to its massive parallelism and low power consumption. However, current CIM designs suffer from large area overhead of small CIM macros and bad programmablity for model execution. This paper proposes a programmable CIM processor with a single large sized CIM macro instead of multiple smaller ones for power efficient computation and a flexible instruction set to support various binary 1-D convolution Neural Network (CNN) models in an easy way. 
% Under the physical limit of CIM macro, the proposed architecture exploits a novel dataflow to improve data transmission that achieves high throughput and low power consumption while provides high flexibility for control and avoids low hardware uttilization. 
Furthermore, the proposed architecture adopts the pooling write-back method to support fused or independent convolution/pooling operations to reduce 35.9\% of latency, and the flexible ping-pong feature SRAM to fit different feature map sizes during layer-by-layer execution.
The design fabricated in TSMC 28nm technology achieves 150.8 GOPS throughput and 885.86 TOPS/W power efficiency at 10 MHz when executing our binary keyword spotting model, which has higher power efficiency and flexibility than previous designs.

% In this paper, we proposes a programmable CIM (Computation-In-Memory) processor with large memory capacity and flexible arithmetic architecture that can widely support various binary 1-D CNN (Convolution Neural Network) models. Under the physical limit of CIM macro, the proposed architecture exploits a novel dataflow to improve data transmissions that achieve high throughput and low power consumption while maintaining high flexibility on controlling and avoiding low hardware utility. PCSNN fabricated in 28nm technology achieves 150.8 GOPS throughput and 885.86 TOPS/W power efficiency at 10 MHz while executing our binary keyword spotting model, performance of which is better than the state-of-the-art keywork spotting CIM hardware.

\end{abstract}

%\begin{IEEEkeywords}
% Spiking neural network, deep learning accelerators
%\end{IEEEkeywords}

\section{Introduction}

%\subsection{Keyword Spotting (KWS)}

% \subsection{Deep Learning Accelerator (DLA)}
Convolution neural network (CNN) is widely used in many artificial intelligence and deep learning applications, such as classification, object detection, and keyword spotting~\cite{KWS2014}. To achieve higher accuracy in different applications, deeper and more complicated network structures have been developed that significantly increases the demands of computation  and memory bandwidth. Thus, speedups with deep learning accelerators (DLAs) are necessary for low power and real-time applications. However, digital DLAs will consume considerable energy consumption and latency on data transmission between PE and memory due to separated computing and storage. Thus, computing-in-memory (CIM) becomes a more viable solution that combines computing and storage together~\cite{CIMservey2021}. The CIM macro uses its crossbar array structure for both weight storage and parallel computation that can achieve massive parallelism for computation and save energy for data transfer.

Previous CIM designs \cite{CIMservey2021, ruiqiguo2019kwscim, dbouk2021kwscim} are dedicated for certain applications or fail to consider programmability to execute models efficiently without complex control. This makes them hard to adapt to different layer execution or incur high control costs. Besides, these designs use multiple small macros to map the model, which is not area and energy efficient due to the large area overhead in small macros and lower available parallelism. 

% Commonly, an arithmetic architecture of a DLA is designed for implementing a certain CNN model. This kind of DLA would have low utilities, or even not work while dealing with other CNN models. To deal with various CNN models, it is necessary to design a general and flexible arithmetic architecture for CNN operation. 

% However, being restricted to physical structure of CIM macros, only weight stationary dataflows can be exploited on CIM-based DLAs. It leads to the CIM-based DLAs having a lower flexibility than the conventional digital DLAs on controlling. Thus, a comprehensive instruction set architecture is necessary for a CIM-based DLA to handle different CNN models. 

To address this issue, we propose a programmable SRAM-based CIM processor for CNN: PSCNN. This design uses a single large CIM macro for high power efficiency up to 886.85 TOPS/W for the keyword spotting application. With this large macro, we propose a comprehensive instruction set architecture that is easily programmable to support various binary 1-D CNN mode executions. Together with this, we adopt 
%a flexible dataflow to increase throughput and reduce power consumption. 
the pooling write-back method to accelerate fused or independent convolution/pooling operations, and the flexible ping-pong feature SRAM to fit various feature map sizes during layer-by-layer execution.
%Furthermore, the proposed architecture adopts the pooling write back method to support fused or independent convolution/pooling operations to reduce 35.9\% latency, and flexible ping-pong feature SRAM to fit different feature map size during layer-by-layer execution.

The rest of the paper is organized as follows. Section II presents the proposed architecture and its data flow. Section III shows the experimental results and comparisons with other designs. Finally, this paper is concluded in Section IV.

\begin{figure}[t]
\centering
\includegraphics[width=0.4\textwidth]{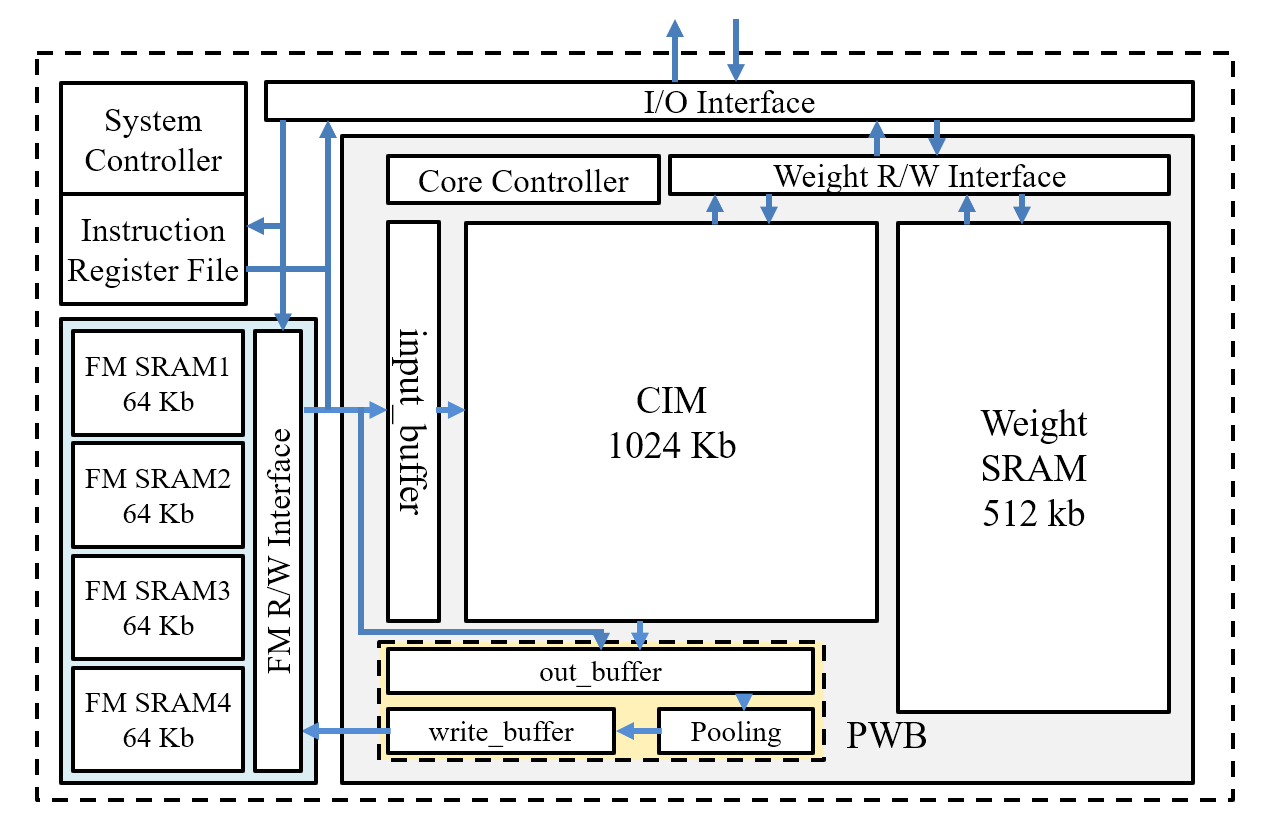}
\caption{The overall architecture of PSCNN.}
\label{Architecture}
\end{figure}

\section{Architecture and Dataflow}
\subsection{Instruction set}
\begin{figure}[t]
\centering
\includegraphics[width=0.47\textwidth]{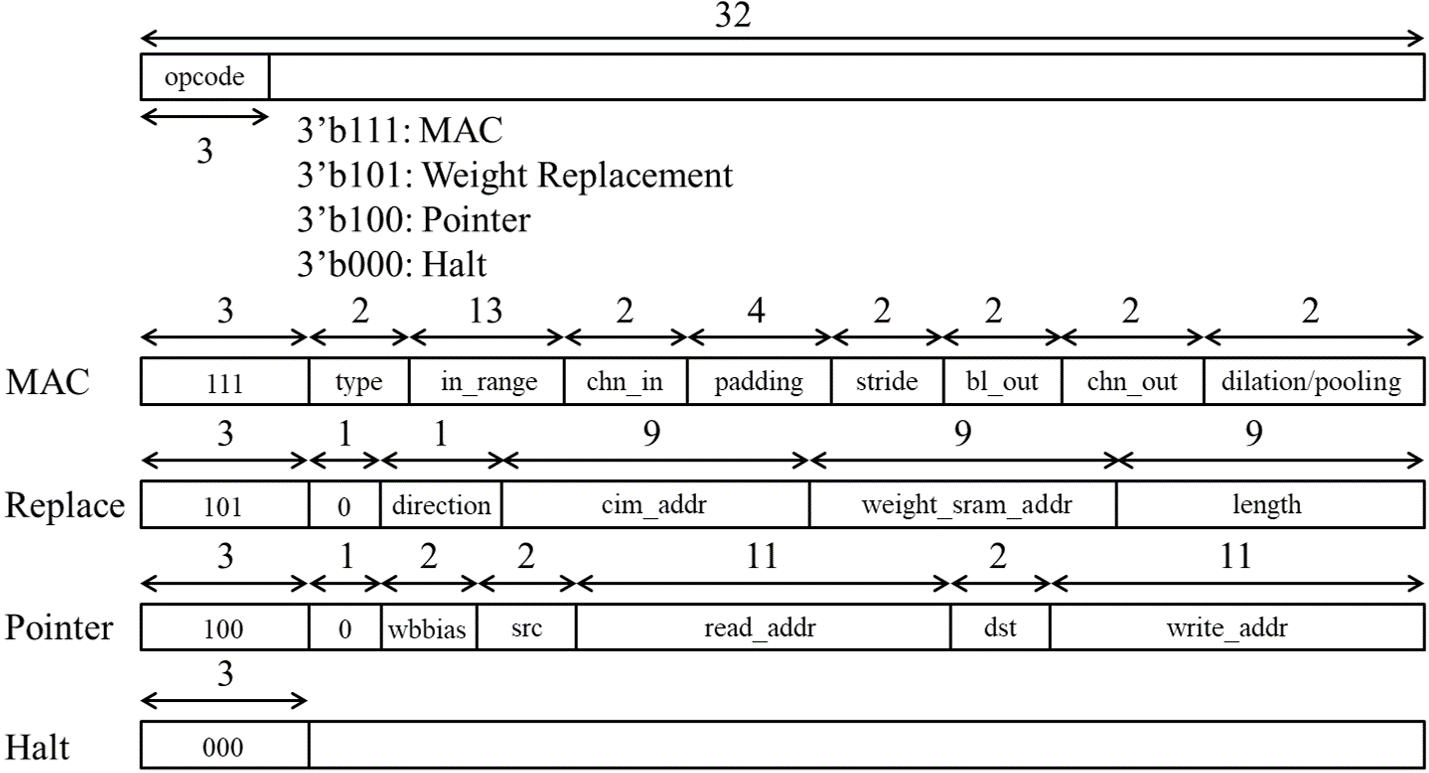}
\caption{Proposed instruction set}
\label{instruction set}
\end{figure}

Fig.~\ref{instruction set} shows the proposed instruction set to program this processor. The instruction set is 32-bit wide and includes four types of instructions: multiply-accumulate (MAC), weight replacement, pointer and halt, depending on the first three bits of the instruction code.

The MAC instruction sets the layer execution type to be convolution or pooling, which is decided by the layer parameters. The weight replacement instruction controls the weight data exchanging between the CIM marco and the weight SRAM if the CIM macro cannot store the whole model. The pointer instruction sets the address of feature map access to dynamically decide the read starting address of the input feature map (IFM) and the write address of the output feature map (OFM) during model execution. The halt instruction stops the controller fetching instructions as long as the model execution is finished. With these instructions, we can easily program CIM without complex control.

%Fig.~\ref{instruction set} shows the proposed instruction set to program this processor. The instruction sets are 32-bit wide and include two types of instruction: external and internal, depending on the MSB of the instruction code. 
%The external type instruction is for the memory access between local memory in PSCNN and external memory. As shown in Fig.~\ref{instruction set}, the second bit of external type instruction determines whether to write data into SRAM/CIM macro, or read data from SRAM/CIM macro, and the next two bits determine which SRAM to be read or written. The starting address and the length of data sequence are determined by the remaining bits. The internal type instruction is for model execution, including MAC, weight replacement and target SRAM setting. With these instructions, we can easily programmable CIM without complex control.

\subsection{Overview}

Fig.~\ref{Architecture} shows the overall architecture of PSCNN that is composed of a system controller, a flexible Ping-Pong SRAM system for I/O feature maps, and one CIM core. The controller extracts the instruction codes from the instruction register file and sends the corresponding control signal to the whole system. The CIM core contains a CIM macro, weight SRAM, a 1024-bit line buffer, and a pooling-write block (PWB). The adopted CIM macro based on \cite{ITRI2021CIM} is a 1Mb SRAM CIM macro and consists of 1024 wordlines, 1024 bitlines, and 128 sense amplifiers (SAs).

% Taking advantage of SRAM CIM being both a PE and a memory unit, the power consumption and latency of data transfer between the PE and memory is largely reduced. Additionally, benefit from the high computing parallelism, the SRAM CIM macro can produce a high throughput.

% \subsection{CIM macro}

\subsection{Single core architecture with a large sized CIM macro}
The small-sized CIM macro is not area and energy efficient. Previous works\cite{ruiqiguo2019kwscim, dbouk2021kwscim} uses multiple small-sized CIM macros for large memory storage to store model parameters for model execution. A small-sized single CIM macro computes only a few multiplications and accumulations (MACs), or a partial sum of one output channel. The partial sums from different CIM macros need to be summed up together additionally by digital circuits. Thus, the output of the CIM macro is not the desired activation output but only intermediate values, which demands high-resolution analog-to-digital converters (ADCs) to sample the output of CIM macros for low truncation error of partial sum, and incurs large area and high power consumption. This situation occurs even for the binary activation network. This implies that the benefits of low precision network no longer exist. 

In contrast, this paper adopts a single core architecture with a large-sized CIM macro for binary activation ternary weight networks. With 1024 wordlines, our CIM macro can compute 1024 MACs on a bitline to generate the final activation output directly without intermediate partial sums and corresponding burdens. Thus, simple SA instead of high resolution ADCs can be used for low power consumption. Besides, the communication network between macros is not necessary due to one CIM core. This high parallelism enables high throughput and allows one channel one bitline mapping for low power consumption.

Besides, when executing a large CNN model, a small-sized CIM macro cannot accommodate the whole model and thus needs to update the weight frequently that will lead to considerable latency and energy consumption. To minimize the impact of weight updating, a large-size SRAM CIM macro is chosen in this work instead of a small-sized one. A large memory storage makes it possible to execute a large model without or with less weight updating need.

\begin{figure}[t]
\centering
\subfigure[Binary-weight mapping]{
\label{Binary-weight}
\includegraphics[height=0.2\textwidth]{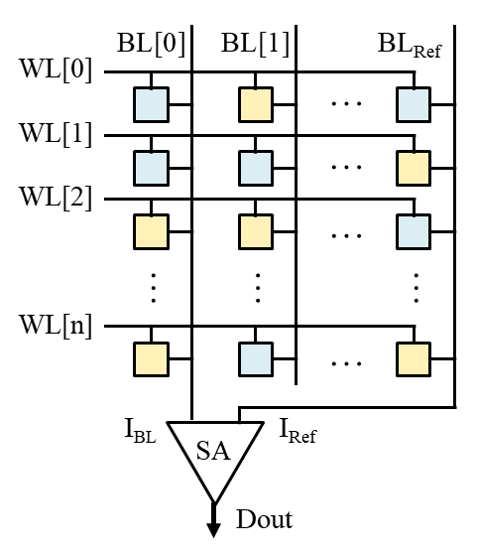}}
\subfigure[Ternary-weight mapping]{
\label{Ternary-weight}
\includegraphics[height=0.2\textwidth]{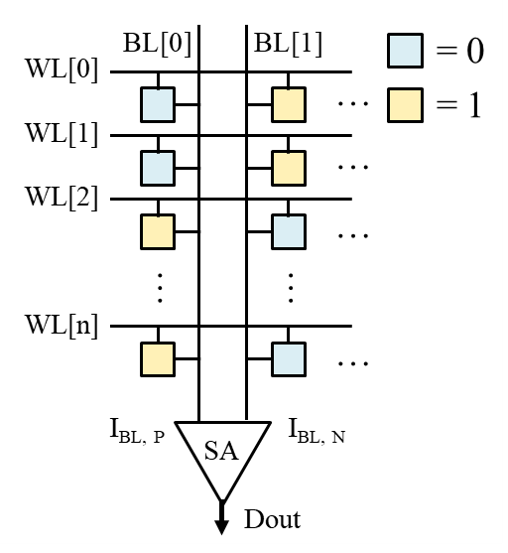}}
\subfigure[Sensing margin comparison]{
\label{sensing_margin}
\includegraphics[height=0.15\textwidth]{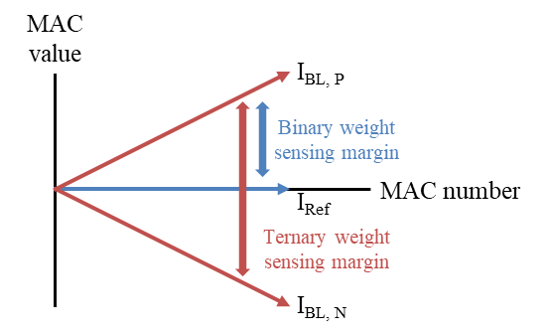}}
\caption{Difference between binary-weight and ternary-weight mapping}
\label{SRAM_cell}
\end{figure}

\subsection{Ternary-weight mapping}
The target model for this design is binary neural network (BNN)~\cite{hubara2016binarized} models using binary precision \{+1, -1\} for weight and \{+1, 0\} for activation. A typical BNN mapping on a CIM macro will use one SRAM cell for one weight as shown in Fig.~\ref{Binary-weight}. After activating the wordlines, the current difference between a bitline and reference bitline is sampled by a SA and regarded as an MAC result. In the ideal situation, the SA plays the role of binarization function as (\ref{eq SA}) in the neural network. 

% In this work, we design the hardware for BNN (binary neural network)~\cite{hubara2016binarized} models using binary precision \{+1, -1\} for weight and \{+1, 0\} for activation. Generally, for a BNN mapping on a CIM macro, each weight occupies one SRAM cell, shown as Fig.~\ref{Binary-weight}. After activating the wordlines, the current difference between a bitline and reference bitline is sampled by a SA (sensing amplifier), regarded as an MAC result. In the ideal situation, the SA plays the role of binarization function (Eq. (1)) in the neural network. 
\begin{equation}
\label{eq SA}
    Dout
    \left\{
             \begin{array}{lr}
             1, & I_{BL} - I_{Ref} \geq 0  \\
             0, & I_{BL} - I_{Ref} < 0
             \end{array}
    \right.
\end{equation}
However, due to the nonideal effect of SA, if the current difference is too small, the actual result will not be detected correctly. 

To prevent the functional failure from SA sensing variation, we replace binary weight mapping (BWM) by ternary weight mapping (TWM)~\cite{ITRI2021CIM}. As shown in Fig.~\ref{Ternary-weight}, different from BWM, each weight of TWM occupies a pair of SRAM cells on two adjacent bitlines. After activating the wordlines, the currents on a bitline pair present the positive popcount and the negative popcount, respectively. An SA compares the currents between the positive one and the negative one instead of comparing with the reference bitline. By this way, TWM requires no additional reference bitline, and the sensing margin is doubled as illustrated in Fig.~\ref{sensing_margin}. Thus, to increase the immunity to variation, we adopt TWM strategy instead of BWM in this work.
\begin{figure}[t]
\centering
\includegraphics[width=0.45\textwidth]{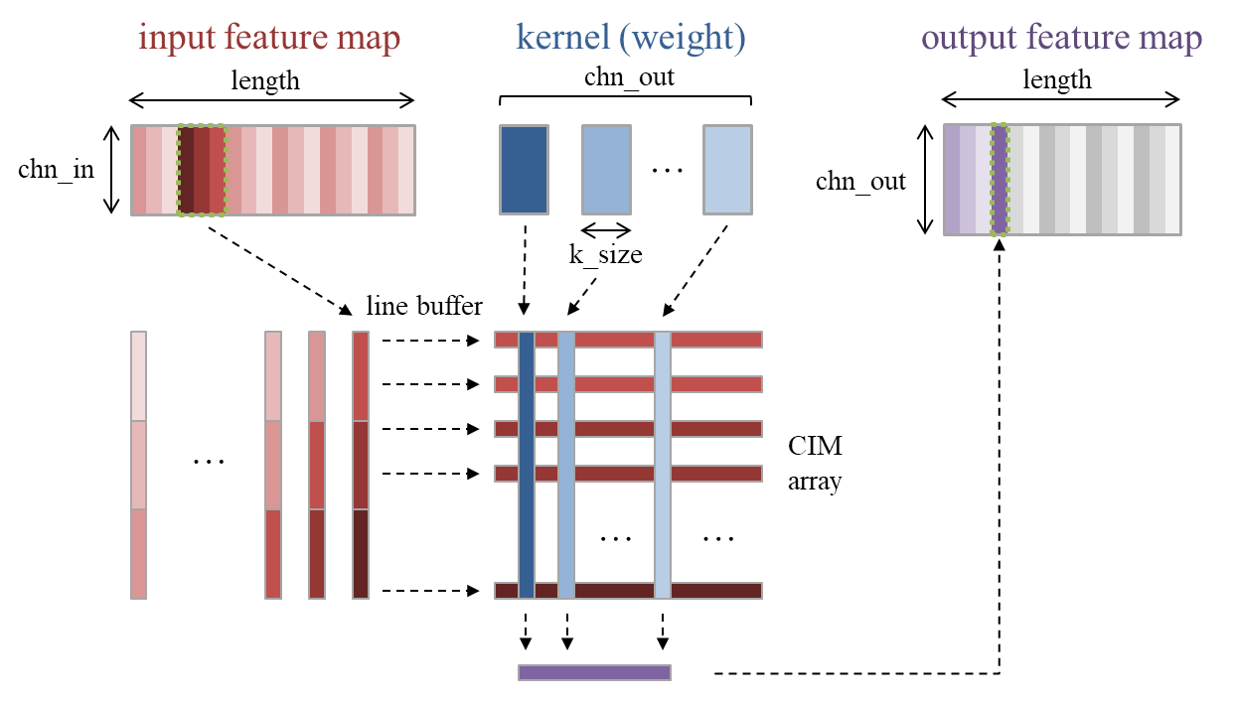}
\caption{Weight Mapping}
\label{Weight_Mapping}
\end{figure}

\subsection{Data flow}
The CIM macro stores weights in the memory and thus uses weight stationary dataflow for computation. Besides, since most of the model weights are available on chip, it is nature to route the macro output directly back to the macro input. Thus, the order of OFM has to be identical to the order of IFM for smooth execution. 

Fig.~\ref{Weight_Mapping} illustrates how to map IFM onto the input line buffer and weights to a crossbar array. The IFM and weights are both grouped based on position indexes, and are arranged in order sequentially. To make sure all outputs are in a correct order, the weights with the same output channel index are placed on the same bitline pair. Convolution layer computations can be done by shifting the IFM downward and activating wordlines alternately. By this mapping method, the order of the OFM can be ensured identical to the IFM, and thus the OFM can be written back directly without data reordering.

% Due to the physical limit of SRAM CIM macro, only weight stationary dataflow can be applied. Although a SRAM CIM macro has a high computation parallelism, a flexible controller and appropriate weight mapping are still necessary to avoid a low hardware utility.

% To execute a CNN model, all weights of each kernel have to be pre-stored in SRAM CIM macros. The order of OFM has to be identical to IFM (input feature map), or OFM needs to be reordered before writing back, which makes some overheads on hardware. Therefore, an appropriate weight mapping is needed. Fig.~\ref{Weight_Mapping} illustrates how to map IFM onto line buffer and to map weights onto a crossbar array. The IFM and weights are both grouped based on position indexes, and are arranged in order sequentially. To make sure all outputs are in a correct order, the weights with the same output channel index are placed on the same bitline pair. Convolution layer computations can be done by shifting the IFM downward and activating wordlines alternately. By this mapping method, the order of the OFM can be ensured identical to the IFM, so the OFM can be written back directly without reordering.

\subsection{Flexible ping-pong feature SRAM}
\begin{figure}[t]
\centering
\subfigure[]{
\label{FM_SRAM(a)}
\includegraphics[width=0.11\textwidth]{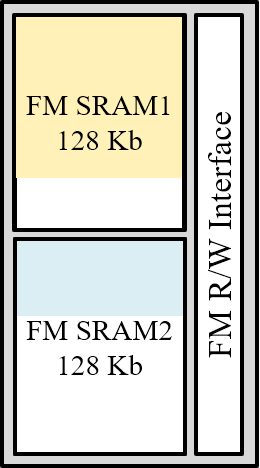}}
\subfigure[]{
\label{FM_SRAM(b)}
\includegraphics[width=0.11\textwidth]{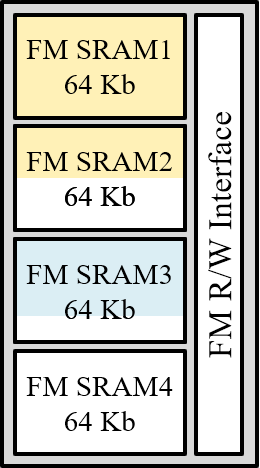}}
\subfigure[]{
\label{FM_SRAM(c)}
\includegraphics[width=0.11\textwidth]{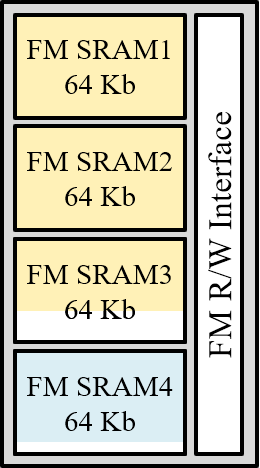}}
\subfigure[]{
\label{FM_SRAM(d)}
\includegraphics[width=0.11\textwidth]{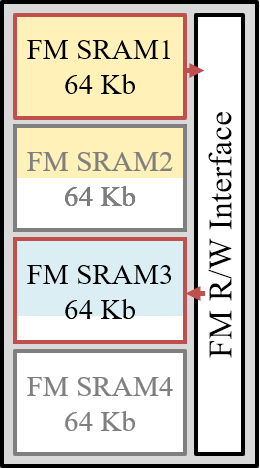}}
\caption{The flexible Ping-Pong feature SRAM system.}
\label{FM_SRAM}
\end{figure}

With the whole model stored on chip, it is nature to execute a model layer by layer and store OFM of each layer in the internal buffer instead of the external buffer to reduce external memory access. Thus, the OFM of the previous layer is the IFM of the next layer, which can be implemented by a ping pong buffer.  Fig.~\ref{FM_SRAM} shows the proposed flexible ping-pong feature SRAM with four 64Kb buffers to store input and output feature maps. This buffer works like a typical ping pong buffer that one buffer will serve as input and another buffer will serve as output for one layer execution and their roles will be switched for the next layer execution. In our design, the read pointer of IFM and the write pointer of OFM can be switched through the instruction assignment of each layer.

However, the conventional ping pong buffer design allocates a fixed size buffer for IFM and OFM, respectively, which could lead to the buffer underutilized as in Fig.~\ref{FM_SRAM} (a) due to the gradually shrinking OFMs by the pooling layers. To reduce such waste, this paper uses four 64Kb single port SRAM macros instead of two larger size ones. Through the instruction assignment, the addresses of IFM and OFM can be determined individually. The memory allocation can be assigned much appropriately, making the processor able to deal with a large-sized feature map, as shown in Fig.~\ref{FM_SRAM}(c). Moreover, only two SRAMs will be either read or written during the convolution calculation. The other two SRAMs can be turned off to save power, as shown in Fig.~\ref{FM_SRAM}(d).

\begin{figure}[t]
\centering
\includegraphics[width=0.40\textwidth]{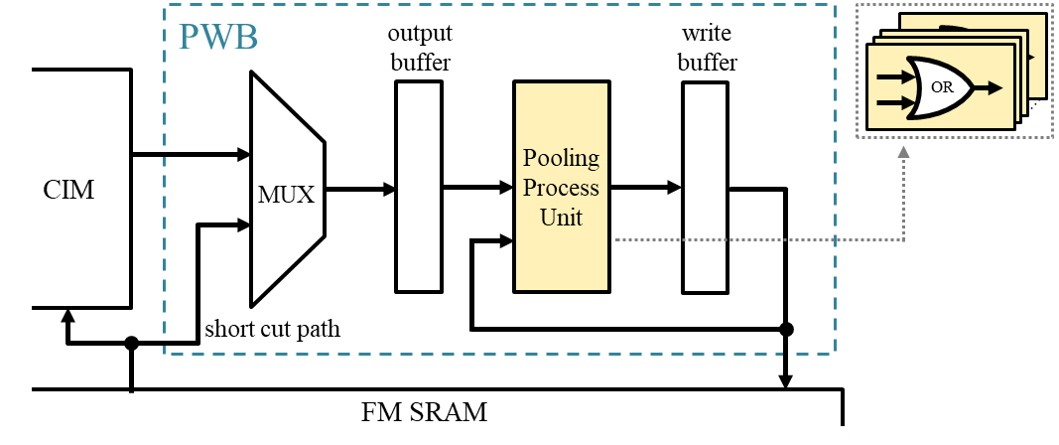}
\caption{Pooling-write block architecure.}
\label{PWB}
\end{figure}

\subsection{Weight SRAM}
Due to the TWM, the effective memory size of CIM macro is 512Kb for weight storage. Although this size is large, the modern model size could easily exceed this limit. Thus, weight updating is inevitable. However, weight updating from DRAM does not enjoy the parallelism of CIM macro and will cause significant latency and power consumption. To reduce latency and power, a 512Kb SRAM is used to store the remaining parameters of a large-sized CNN model. Through the instruction assignment, the parameters of the SRAM CIM macro can be replaced with the weight data stored in the weight SRAM if needed. With additional memory storage, PSCNN can handle large-sized CNN models in high speed and low power.

% Due to the TWM, the effective memory size of CIM macro is 512Kb. Even though a large-sized SRAM CIM macro has been adopted, it is still possible that it can’t store the whole parameter of a deep CNN model. Thus, weight updating is inevitable. The weight data transmission from DRAM will cause significant latency and power consumption. Benefit from the high computation parallelism of the SRAM CIM macro, the latency contributed from convolution is much less than what data transmission does. The time delay caused by weight data transmission would become the bottleneck. To reduce latency and power, a 512Kb SRAM is used to store the remaining parameter of a large-sized CNN model. Compared to accessing weight data from DRAM, latency and power considerably decreases to access data from local SRAM. Through the instruction assignment, the parameter on SRAM CIM macro can be replaced with the weight data stored in the weight SRAM if needed. With additional memory storage, PSCNN can handle large-sized CNN models in high speed and low power.

\subsection{Pooling-Write Block (PWB)}
A typical model usually has a pooling layer following a convolution layer. Separately executing these two layers will need to fetch convolution data from SRAM again for pooling, which consumes additional latency and power. 

\begin{figure}[t]
\centering
\includegraphics[width=0.45\textwidth]{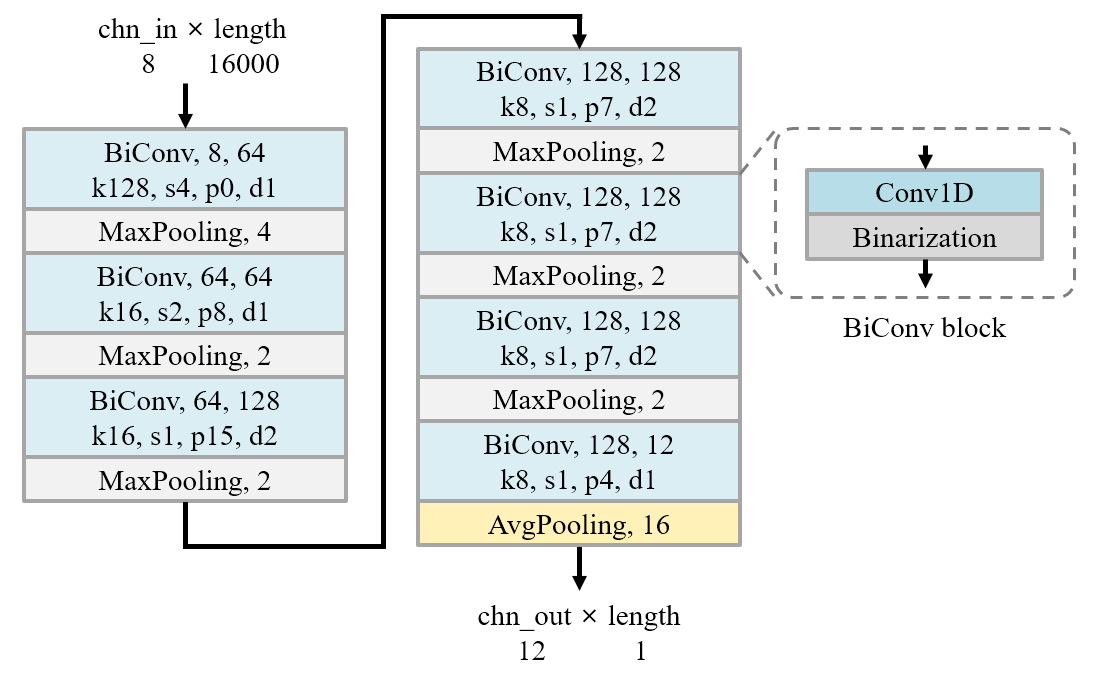}
\caption{The configuration of our binary keyword spotting model.}
\label{kws_network}
\end{figure}

To avoid this, this paper proposes PWB that adds pooling operations after the output of SRAM CIM macro to skip the redundant data transfers because the OFM of a convolution layer generated by CIM macro in every cycle is in order. Moreover, by this way, the pipelined convolution and pooling can avoid stalls and improve hardware utilization. 
By running the test model shown in Fig.~\ref{kws_network}, with the PWB supporting fused convolution and pooling operation, the execution latency is reduced by 35.9\%.
However, simply adding pooling units after the macro will limit its flexibility. In some cases, convolution and pooling need to be executed individually. In such a case, a shortcut path can be activated to bypass the SRAM CIM macro, as illustrated in Fig~\ref{PWB}. In summary, by the instruction assignment, the convolution layer and pooling layer can be executed independently, or be fused together to reduce latency depending on the situation.

\section{Experimental Results}

\subsection{Network simulation result}
Fig.~\ref{kws_network} shows our binary keyword spotting model with 652Kb model size. The performance of the model is evaluated on the Google speech commands dataset (GSCD)~\cite{warden2018speech}. The input sound data is 1 second length at 16kHz sampling rate, and quantized into 8 bits fixed point. The network detects 12 classes of keywords including unknown and silence from the input sound data. The inference accuracy is 92.53\%. 

%\begin{figure}[t]
%\centering
%\includegraphics[width=0.4\textwidth]{Speedup.png}
%\caption{Latency comparison between data transmission elimination methods.}
%\label{speedup}
%\end{figure}

\begin{table}[t]
\caption{Performance summary and comparison with other designs.}
\begin{tabular}{|c|c|c|c|}
\hline
                        & Think-IM~\cite{ruiqiguo2019kwscim} & KeyRAM~\cite{dbouk2021kwscim} & This work \\ \hline
Technology              & 65nm & 65nm & 28nm \\ \hline
Algorithm               & RNN & RAM & BNN \\ \hline
Dataset                 & GSCD & GSCD & GSCD \\ \hline
\# of Classes           & 10 & 7 & 12 \\ \hline
\begin{tabular}[c]{@{}c@{}}Test Accuracy\\ (\%)\end{tabular} & 90.2 & 90.38 & 92.53 \\ \hline
Architecture            & Digital \& CIM & Digtal \& CIM & CIM only \\ \hline
Precision (bits)        & binary & 8-b \& 4-b & binary \\ \hline
Reconfigurable          & no & no & yes \\ \hline
CIM type                & 6T SRAM & 6T SRAM & 10T SRAM \\ \hline
CIM array size          & 16$\times$64$\times$64 & 2$\times$512$\times$256 & 1$\times$1024$\times$1024 \\ \hline
\begin{tabular}[c]{@{}c@{}}On-Chip Memory\\ w/o CIM (Kb)\end{tabular} & 80 & 96 & 768 \\ \hline
Voltage (V)             & 0.9 & 1 & 0.9 \\ \hline
\begin{tabular}[c]{@{}c@{}}Frequency\\ (MHz)\end{tabular} & 75 & 1000 & 10 \\ \hline
\begin{tabular}[c]{@{}c@{}}Energy per\\ inference ($\mu$J)\end{tabular} & 3.36 & 0.44  & 0.399 \\ \hline
\begin{tabular}[c]{@{}c@{}}Latency per\\ inference ($\mu$s)\end{tabular} & 127.3 & 39.9 & 2320 \\ \hline
\begin{tabular}[c]{@{}c@{}}\# of MACs \\ per inference\end{tabular} & - & 200k & 350M \\ \hline
\begin{tabular}[c]{@{}c@{}}\# of Parameterr\\ (Kb)\end{tabular} & - & 91 & 652 \\ \hline
\begin{tabular}[c]{@{}c@{}}Throughput\\ (GOPS)\end{tabular} & - & 5.5 & 150.8 \\ \hline
\begin{tabular}[c]{@{}c@{}}Power efficiency\\ (TOPS/W)\end{tabular} & 11.7 & 0.91 & 885.86 \\ \hline
\end{tabular}
\label{comparison with other hardware}
\end{table}

\subsection{Implementation results and comparison}
The proposed design has been implemented on TSMC 28nm CMOS process. It can achieve 150.8 GOPS throughput and 885.86 TOPS/W power efficiency at 10 MHz with 23.4K gate counts (8,485.34 $um^{2}$ in area), one 1Mb CIM macro, four 64Kb SRAM macros, and one 512Kb SRAM macro. 

%Fig.~\ref{speedup} shows the ablation study of the proposed method to reduce external data access evaluated on PSCNN running the keyword spotting model. The baseline is the one without any proposed method. For the test model, its 652Kb model size exceeds the size of the CIM macro. The remaining 140Kb parameters are stored in weight SRAM, and being used to update the weight on CIM macro when needed. Assume all the weights have been pre-load in the CIM macro and the weight SRAM. Adopting the flexible Ping-Pong SRAM and the PWB achieves 1.39× and 1.56× individual speedup, and 2.17× overall speedup. 

The performance is evaluated with PSCNN running the keyword spotting model shown in Fig.~\ref{kws_network}. For the test model, its 652Kb model size exceeds the size of the CIM macro. The remaining 140Kb parameters are stored in weight SRAM, and are used to update the weight of CIM macro when needed. Assume all weights have been preload in the CIM macro and the weight SRAM. Table.~\ref{comparison with other hardware} shows the performance summary and comparison with other CIM-based designs for keyword spotting. \cite{dbouk2021kwscim} has lower throughput and power efficiency due to their high precision on weight and activation, which needs high resolution ADCs. ~\cite{ruiqiguo2019kwscim} adopts tile-based architecture, high resolution ADCs, and additional adder trees to deal with multibit partial sum, which leads to lower power efficiency and higher energy consumption. Both ~\cite{dbouk2021kwscim} and~\cite{ruiqiguo2019kwscim} design dedicated dataflows for their own algorithms, which cannot be reconfigured for other models. Compared to other designs, this design has higher power efficiency and high flexibility for model execution.

% The simulation baseline is the architecture of PSCNN without anny transmission elimination method. We implement our binary keyword spotting model on our PSCNN. Due to the 652Kb model size and the TWM, our CIM macro can only accommodate 512Kb parameters. The remaining 140Kb parameters are stored in weight SRAM, and being used to update the weight on CIM macro when needed. Assume all the weights have been pre-load in the CIM macro and the weight SRAM. Adopting the flexible Ping-Pong SRAM and the PWB achieves 1.39× and 1.56× individual speedup, and 2.17× overall speedup, shown in Fig.~\ref{speedup}. PSCNN fabricated in 28nm technology achieves 150.8 GOPS throughput and 885.86 TOPS/W power efficiency at 10 MHz with 23.4K gate counts (8,485.34 $um^{2}$ in area), one 1Mb CIM macro, four 64Kb SRAM macros, and one 512Kb SRAM macros. 

% Table.~\ref{comparison with other hardware} shows the performance summary and comparison with other keyword spotting CIM-based DLA designs. Compared to ~\cite{dbouk2021kwscim}, they have lower throughput and power efficiency due to their high precision on weight and activation, which needs high resolution ADCs. ~\cite{ruiqiguo2019kwscim} adopts tile-based architecture, high resolution ADCs and additional adder trees are utilized to deal with multi-bit partial sum, whose hardware thus have lower power efficiency and higher energy consumption. Both of~\cite{dbouk2021kwscim} and~\cite{ruiqiguo2019kwscim} are design a dedicated dataflow for their own algorithms, they cannot be reconfigured for other models.

\section{Conclusion}
In this paper, we adopt the single large CIM core architecture to eliminate digital hardware overheads and to avoid the truncation error issue. Furthermore, we use ternary weight mapping instead of binary weight mapping to increase the immunity to SA variation. The proposed accelerator is easily programmed with a flexible instruction set to support various binary 1-D CNN models. In addition, the proposed architecture further reduces the latency by the pooling write back method and reduces the memory size requirement with a flexible ping-pong feature SRAM. The proposed design implemented with TSMC 28nm CMOS technology achieves 150.8 GOPS throughput and 885.86 TOPS/W power efficiency at 10 MHz when executing our binary keyword spotting model, which has better power efficiency than the the state-of-the-art designs.

\section*{Acknowledgment}
This work was supported by the Ministry of Science and Technology, Taiwan, under Grant 109-2634-F-009-022, 109-2639-E-009-001, 110-2221-E-A49-148-MY3 and 110-2622-8-009-018-SB, and TSMC.

\bibliographystyle{IEEEtran}
\bibliography{citation.bib}
\end{document}